\documentclass[a4paper, 12pt, openany, oneside]{article}
\usepackage[cp1251]{inputenc}
\usepackage[english, russian]{babel}
\usepackage{amsmath, latexsym,  amssymb, bm, graphics, eucal, amsfonts, array}
\usepackage[dvips]{graphicx}
\makeatletter

\newcommand{\rot}{\mathop{\rm rot\,}}
\makeatother
\usepackage{indentfirst}
\makeatother

\begin{document}
\renewcommand{\refname}{\begin{center} \bf References\end{center}}
\thispagestyle{empty}
\renewcommand{\abstractname}{Abstract}
\renewcommand{\contentsname}{Contents}
 \large
\newcommand{\mc}[1]{\mathcal{#1}}
\newcommand{\E}{\mc{E}}

 \begin{center}
\bf NONLINEAR LONGITUDINAL CURRENT  GENERATED BY  N  TRANSVERSAL
ELECTROMAGNETIC WAVES IN THE QUANTUM PLASMA
\end{center}\bigskip

\begin{center}
\bf A. V. Latyshev\footnote{avlatyshev$@$mail.ru} and
V. I. Askerova\footnote{vera$ \_ $askerova$@$mail.ru}
\end{center}\medskip
\bigskip

\begin{center}
{\it Faculty of Physics and Mathematics,\\ Moscow State Regional
University, 105005,\\ Moscow, Radio str., 10-A}
\end{center}\medskip

\begin{abstract}

Quantum plasma with arbitrary degree of degeneration of electronic gas is consi\-de\-red.
In plasma N (N>2) external electromagnetic waves are propagated. It is required to
find the response of plasma to these waves. From kinetic Wigner equation for quantum
collisionless plasmas distribution function in square-law approach on quantities of
vector potentials of N electric fields is received. The formula for electric
current calculation is deduced at arbitrary temperature, i.e. at arbitrary
degree of degeneration of electronic gas. It is shown, that the nonlinearity
account leads to occurrence of the longitudinal electric current directed
along a wave vector. This longitudinal current is orthogonal to the known
transversal current received at the linear analysis. The case of small values
of wave number is considered. It is shown, that in case of small values of wave
numbers the longitudinal current in quantum plasma coincides with a longitudinal
current in classical plasma.

{\bf Key words:} quantum plasma, Wigner equation, Fermi---Dirac distribution,
longi\-tu\-dinal and transversal electric current, nonlinear analysis.

PACS numbers: 03.65.-w Quantum mechanics, 05.20.Dd Kinetic theory,
52.25.Dg Plasma kinetic equations.
\end{abstract}

\tableofcontents
\setcounter{secnumdepth}{4}

\begin{center}
\item{}\section{Introduction}
\end{center}

In the present work formulas for electric current calculation in quantum collisionless
Fermi---Dirac plasma are deduced. At the decision of the kinetic Wigner equation
describing behavior of plasma, we consider both in expansion of  distribution
function, and in expansion of Wigner integral the quantities proportional to
squares of vector potentials of external electric fields and their products.
In such nonlinear approach it appears, that the electric current has two nonzero
components. One component of an electric current it is directed along vector
potentials of electromagnetic fields. These components of an electric current
precisely same, as well as in the linear analysis. It is a "transversal"\, current.
Those, in linear approach we receive known expression of a transversal electric current.
The second nonzero an electric current component has the second order of smallness
concerning quantities intensity of electric fields. The second electric current
component is directed along a wave vector. This current is orthogonal to the
first a component. It is a "longitudinal"\, current. Occurrence of a longitudinal
current comes to light the spent nonlinear analysis of interaction of electromagnetic
fields with plasma.

Nonlinear effects in plasma are studied already long time \cite{Gins}--\cite{Lat1}.

In works \cite{Gins} and \cite{Zyt} nonlinear effects in plasma are studied.
In work \cite{Zyt} nonlinear current was used, in particular, in probability
questions decay processes. We will notice, that in work \cite{Zyt2} it is
underlined existence of nonlinear current along a wave vector (see the
formula (2.9) from  \cite{Zyt2}).

In experimental work \cite{Akhmediev} the contribution normal field components
in a nonlinear superficial current in a signal of the second harmonic is found out.
In works \cite{Urupin1, Urupin2}
generation of a nonlinear superficial current was studied at interaction of a laser
impulse with metal.

Quantum plasma was studied in works \cite{Lat1}--\cite{Lat9}.
Collisional quantum plasma has started to be studied in work of Mermin \cite{Mermin}.
Then quantum collision plasma was studied in our works \cite{Lat2}--\cite{Lat5}.
In works \cite{Lat7} -- \cite{Lat9} generating of a longitudinal current by a
transversal electromagnetic field in Fermi---Dirac classical and quantum plasma
\cite{Lat7} and into degenerate plasma \cite{Lat9}. We will specify in a number of
works on plasma, including to the quantum. These are works \cite{Andres}--\cite{Manf}.

In this paper formulas for electric current calculation into quantum colli\-sion\-less
plasma are deduced at any temperature, at any degree of degeneration of electronic gas.

\begin{center}
\item{}\section{The Wigner equation}
\end{center}

Let us demonstrate, that in case of the quantum plasma described by kinetic Wigner
equation, the longitudinal current is generated, and we will calculate its density.
It was specified in existence of this current more half a century ago \cite{Zyt2}.

Let us  turn to the consideration of the fact, that the quantum plasma is in N external
electromagnetic fields with the vector potentials representing running harmonious waves
$$
{\bf A}_j({\bf r},t)={\bf A}_{0j}e^{i({\bf k}_j{\bf r}-\omega_jt)}
\quad (j=1,2,\cdots,N).
$$

Corresponding electric and magnetic fields
$$
{\bf E}_j={\bf E}_{0j}e^{i({\bf k_jr}-\omega_j t)}, \qquad
{\bf H}_j={\bf H}_{0j}e^{i({\bf k_jr}-\omega_j t)}, \quad (j=1,2,\cdots,N)
$$
are connected with vector potentials equalities
$$
{\bf E}_j=-\dfrac{1}{c}\dfrac{\partial {\bf A}_j}{\partial t}=
\dfrac{i\omega_j}{c}{\bf A}_j,\qquad
{\bf H}_j=\rot {\bf A}_j \qquad (j=1,2,\cdots,N).
$$

It is necessary to specify, that vector potential of an electromagnetic field
${\bf A}_j({\bf r},t)$  is orthogonal to a wave vector ${\bf k}_j$, т.е.
$$
{\bf k}_j\cdot {\bf A}_j({\bf r},t)=0,\qquad j=1,2,\cdots,N.
$$
It means that the wave vector ${\bf k}_j$ is orthogonal
to electric and magnetic fields
$$
{\bf k}_j\cdot {\bf E}_j({\bf r},t)=
{\bf k}_j\cdot {\bf H}_j({\bf r},t)=0,\qquad j=1,2,\cdots,N.
$$

We take the Wigner equation describing behavior of quantum collisionless plasma
$$
\dfrac{\partial f}{\partial t}+\mathbf{v}\dfrac{\partial f}{\partial
\mathbf{r}}+W[f]=0
\eqno{(1.1)}
$$
with  nonlinear integral of Wigner
$$
W[f]= {\bf p}\sum\limits_{j=1}^{N}\dfrac{ie{\bf A}_j}{mc\hbar}
\Big[f\Big({\bf r},{\bf p}+\dfrac{\hbar{\bf k}_j}{2},t\Big)-
f\Big({\bf r},{\bf p}-\dfrac{\hbar{\bf k}_j}{2},t\Big)\Big]-
$$
$$
-\dfrac{ie^2}{2mc^2\hbar}\sum\limits_{j=1}^{N}{\bf A}_j^2
\Big[f\Big({\bf r},{\bf p+{\hbar k}}_j,t\Big)-
f\Big({\bf r},{\bf p-{\hbar k}}_j,t\Big)\Big]-
$$
$$
-\dfrac{2ie^2}{2mc^2\hbar}\sum\limits_{\substack{s,j=1\\j<s}}^{N}{\bf A}_s{\bf A}_j
\Big[f\Big({\bf r},{\bf p}+\hbar\dfrac{{\bf k}_s+{\bf k}_j}{2},t\Big)-
f\Big({\bf r},{\bf p}-\hbar\dfrac{{\bf k}_s+{\bf k}_j}{2},t\Big)\Big].
$$\medskip

This integral of Wigner is constructed by analogy to integral of Wigner
deduced in work \cite{Lat1} with use of one vector potential.

In the equation (1.1) $f$ is the analogue of quantum function of distribution
electrons plasmas (so-called function of Wigner), $c$ is the velocity of light,
${\bf p}=m{\bf v}$ is the electrons momentum, ${\bf v}$ is the electrons velocity.

Lower local equilibrium distribution of Fermi---Dirac is required to us,
$f^{(0)}=f_{eq}({\bf r},v)$ (eq$\equiv$ equi\-lib\-rium),
$$
f_{eq}({\bf r},v)=\Big[1+\exp\dfrac{\E-\mu({\bf
r})}{k_BT}\Big]^{-1},
$$
$\E={mv^2}/{2}$ is the electron energy, $\mu$ is the chemical potential of
electronic gas, $k_B$ is the Boltzmann constant, $T$ is the plasma temperature,
$v_T$ is the heat electron velocity,
$$
v_T=\sqrt{\dfrac{2k_BT}{m}},\qquad k_BT=\E_T=\dfrac{mv_T^2}{2},
$$
$\E_T$ is the heat kinetic electron energy.

In quantum plasma velocity of electrons is connected with its momentum and vector
potentials of electromagnetic fields equality
$$
{\bf v}=\dfrac{{\bf p}}{m}-\dfrac{e}{cm}({\bf A}_1+{\bf A}_2+\cdots+{\bf A}_N).
$$

If we introduce the dimensionless momentum of the electrons
${\bf P}={{\bf p}}/{p_T}$, where  $p_T=mv_T$ is the heat  electron momentum,
then the preceding equation can be written as
$$
{\bf v}=v_T\Big({\bf P}-\dfrac{e}{cp_T}({\bf A}_1+{\bf
A}_2+\cdots+{\bf A}_N)\Big).
$$

We need absolute distribution of Fermi---Dirac
$f_0(p)$,
$$
f_0(p)=\Big[1+\exp\dfrac{p^2/2m-\mu}{k_BT}\Big]^{-1}=
\Big[1+\exp\Big(\dfrac{p^2}{p_T^2}-\alpha\Big)\Big]^{-1}=$$$$=
\dfrac{1}{1+e^{P^2-\alpha}}=f_0(P).
$$

Here $\alpha=\mu/(k_BT)$ is  the chemical potential of electronic gas.

We pass to dimensionless momentum in equation (1.1) and in Wigner' integral.
We obtain the equation
$$
\dfrac{\partial f}{\partial t}+v_T{\bf P}
\dfrac{\partial f}{\partial {\bf r}}+
 \dfrac{ie v_T}{c\hbar}\sum\limits_{j=1}^{N}({\bf P}{\bf A}_j)
\Big[f\big({\bf r},{\bf P}+\dfrac{{\bf q}_j}{2},t\big)-
f\big({\bf r},{\bf P}-\dfrac{{\bf q}_j}{2},t\Big)\big]-
$$
$$
-\dfrac{ie^2}{2mc^2\hbar}\sum\limits_{j=1}^{N}{\bf A}_j^2
\Big[f({\bf r},{\bf P+{q}}_j,t)-
f({\bf r},{\bf P}-{\bf q}_j,t)\Big]-
$$
$$
-\dfrac{2ie^2}{2mc^2\hbar}\sum\limits_{\substack{s,j=1\\j<s}}^{N}{\bf A}_s{\bf A}_j
\Big[f\big({\bf r},{\bf P}+\dfrac{{\bf q}_s+{\bf q}_j}{2},t\big)-
f\big({\bf r},{\bf P}-\dfrac{{\bf q}_s+{\bf q}_j}{2},t\big)\Big]=0.
\eqno{(1.2)}
$$

Here ${\bf q}_j$ is the dimensionless wave numbers,
${\bf q}_j={\bf k}_j/{k_T}$, $k_T$ is the heat wave number, $k_T=mv_T/\hbar$.

For definiteness we will consider, that wave vectors $N$ of fields are directed along an
axis $x$ and electromagnetic fields are directed along an axis  $y$,i.e.
$$
{\bf A}_j=A_j(x,t)(0,1,0),\qquad A_j(x,t)=A_{0j}e^{i(k_jx-\omega_jt)},
$$
$$
{\bf k}_j=k_j(1,0,0), \quad {\bf E}_j=E_{j}(x,t)(0,1,0),\quad
E_j(x,t)=E_{0j}e^{i(k_jx-\omega_j t)}.
$$

Now the equation (1.2) is somewhat simplified
$$
\dfrac{\partial f}{\partial t}+v_TP_x
\dfrac{\partial f}{\partial x}+
\dfrac{ie v_T}{c\hbar}\sum\limits_{j=1}^{N}(P_y A_j)
\Big[f\big(x,{\bf P}+\dfrac{{\bf q}_j}{2},t\big)-
f\big(x,{\bf P}-\dfrac{{\bf q}_j}{2},t\Big)\big]-
$$
$$
-\dfrac{ie^2}{2mc^2\hbar}\sum\limits_{j=1}^{N}A_j^2
\Big[f(x,{\bf P+{q}}_j,t)-
f(x,{\bf P}-{\bf q}_j,t)\Big]-
$$
$$
-\dfrac{2ie^2}{2mc^2\hbar}\sum\limits_{\substack{s,j=1\\j<s}}^{N}A_sA_j
\Big[f\big(x,{\bf P}+\dfrac{{\bf q}_s+{\bf q}_j}{2},t\big)-
f\big(x,{\bf P}-\dfrac{{\bf q}_s+{\bf q}_j}{2},t\big)\Big]=0.
\eqno{(1.2')}
$$

We will search the solution of equation (1.2) in the form
$$
f=f_0(P)+f_1(x,{\bf P},t)+f_2(x,{\bf P},t).
\eqno{(1.3)}
$$
Here
$$
f_1(x,{\bf P},t)=A_1(x,t)\varphi_1({\bf P})+A_2(x,t)\varphi_2({\bf P})+\cdots+
A_N(x,t)\varphi_N({\bf P})=
$$
$$
=\sum\limits_{j=1}^{N} A_j(x,t)\varphi_j({\bf P}),
\eqno{(1.4)}
$$
where
$$
A_{j}(x,t)\sim e^{i(k_{j}x-\omega_{j}t)},
$$
$$
f_2(x,{\bf P},t)=\sum\limits_{b=1}^{N}{A}^{2}_b(x,t)\psi_{b}({\bf P})+
\sum\limits_{\substack{s,j=1\\j<s}}^{N}{A}_s(x,t){A}_j(x,t)\xi_{j,s}({\bf P}),
\eqno{(1.5)}
$$
where
$$
A_{b}(x,t)\sim e^{2i(k_{b}x-\omega_{b}t)}, A_{s}(x,t)A_{j}(x,t)
\sim e^{i[(k_{s}+k_{j})x-(\omega_{s}+\omega_{j})t]}.
$$
In equalities (1.4) and (1.5) new unknown functions are introduced
$$
\varphi_j=\varphi_j({\bf P}),\qquad \xi_{s,j}=\xi_{s,j}({\bf P}),\qquad
\psi_{b}=\psi_{b}({\bf P}),\qquad j,s,b=1,2,\cdots, N.
$$

\begin{center}
\item{}\section{The solution of Wigner equation in first approximation}
\end{center}

In this equation exist 2N parameters of dimension of length
$\lambda_j={v_T}/{\omega_j}$ ($v_T$ is the heat electron velocity) and
$l_j={1}/{k_j}$. We shall believe, that on lengths $\lambda_j$, so and on lengths
$l_j$ energy variable of electrons under acting correspond electric field $A_j$ is
much less than heat energy of electrons $k_BT$ ($k_B$ is the Boltzmann constant, $T$ is
the temperature of plasma), i.e. we shall consider small parameters
$$
\alpha_j=\dfrac{{\left|eA_j\right|v_T}}{c k_BT}\qquad (j=1,2,\cdots,N)
$$
and
$$
\beta_j=\dfrac{{\left|eA_j\right|\omega_j}}{k_j k_BTc}\qquad (j=1,2,\cdots,N).
$$

If to use communication of vector potentials
electromagnetic fields with strengths of corresponding
electric fields, injected small parameters are expres\-sed
following equalities
$$
\alpha_j=\dfrac{{\left|eE_j\right|v_T}}{\omega_j k_BT}\qquad (j=1,2,\cdots,N)
$$
and
$$
\beta_j=\dfrac{{\left|eE_j\right|}}{k_j k_BT}\qquad (j=1,2,\cdots,N).
$$

We will work with a method consecutive approximations, considering, that
$$
\alpha_j\ll 1 \qquad (j=1,2,\cdots,N)
$$
and
$$
\beta_j\ll 1  \qquad (j=1,2,\cdots,N).
$$

In the first approximation we search the solution of Wigner equation in the form
$$
f=f^{(1)}=f_0(P)+f_1,
\eqno{(2.1)}
$$
where $f_1$ is the linear combination of vector potentials (1.4).

The Wigner equation (1.2) in linear approximation on quantities vector potentials
has the following form
$$
\dfrac{\partial f}{\partial t}+v_TP_x\dfrac{\partial f}{\partial x}+
\dfrac{iev_T}{c\hbar}\sum\limits_{j=1}^{N}(P_y A_j)\Big[f(x,{\bf P}+\dfrac{{\bf q}_j}{2},t)-
f(x,{\bf P}-\dfrac{{\bf q}_j}{2},t)\Big]=0.
\eqno{(2.2)}
$$ \bigskip

By means of a method of small parameter, we will substitute in the first two members
of equation (2.2) $f=f_1$, and in the third member of equation  $f=f_0$. We receive the
following equation
$$
\dfrac{\partial f_1}{\partial t}+v_TP_x\dfrac{\partial f_1}{\partial x}+
\dfrac{iev_T}{c\hbar}\sum\limits_{j=1}^{N}(P_y A_j)\Big[f_0({\bf P}+\dfrac{{\bf q}_j}{2})-
f_0({\bf P}-\dfrac{{\bf q}_j}{2})\Big]=0.
\eqno{(2.2')}
$$ \bigskip

Here
$$
f_0\Big({\bf P}\pm \dfrac{{\bf q}_j}{2}\Big)=\left\{1+
\exp\Big[\Big({\bf P}\pm \dfrac{{\bf q}_j}{2}\Big)^2-\alpha\Big]\right\}^{-1}=
$$
$$
=\left\{1+ \exp\Big[\Big(P_x\pm \dfrac{q_j}{2}\Big)^2+P_y^2+P_z^2-\alpha\Big]\right\}^{-1}
$$
and the dimensionless parameters are introduced
$$
q_j=\dfrac{k_j}{k_T},\qquad j=1,2,\cdots,N,
$$
$q_j$ is the dimensionless wave number, $k_T =\dfrac {mv_T} {\hbar} $ is the heat wave number.
We introduce also the dimensionless oscillation frequency of vector potential of
electromagnetic field ${\bf A}_j$, $ \Omega_j=\dfrac{\omega_j}{k_Tv_T}$.

At substitution (2.1) and (1.4) in the equation (2.2) we receive the following equation
$$
A_1\varphi_1(q_1P_x-\Omega_1)+\cdots+A_N\varphi_N(q_NP_x-\Omega_N)=
$$
$$
=-\dfrac{ev_T}{c\hbar k_Tv_T}\Big\{P_yA_1\Big[f_0\Big({\bf P}+\dfrac{{\bf q}_1}{2}\Big)-
f_0\Big({\bf P}-\dfrac{{\bf q}_1}{2}\Big)\Big]+\cdots+
$$
$$
+P_yA_N\Big[f_0\Big({\bf P}+\dfrac{{\bf q}_N}{2}\Big)-
f_0\Big({\bf P}-\dfrac{{\bf q}_N}{2}\Big)\Big]\Big\}.
$$
The last equation breaks up on the equations $N$ the equations
$$
(q_jP_x-\Omega_j)A_j\varphi_j=-
\dfrac{e(P_yA_j)}{c\hbar k_T}\Big[f_0({\bf P}+\dfrac{{\bf q}_j}{2})-
f_0({\bf P}-\dfrac{{\bf q}_j}{2})\Big],\quad j=1,2,\cdots,N,
$$
from which we receive
$$
A_j\varphi_j=-\dfrac{eP_yA_j}{c\hbar k_T}\dfrac{f_0({\bf P}+\dfrac{{\bf q}_j}{2})-
f_0({\bf P}-\dfrac{{\bf q}_j}{2})}{q_jP_x-\Omega_j},\quad j=1,2,\cdots,N.
$$

Thus Wigner's function is as a first approximation  is constructed
$$
f_1=-\dfrac{e}{c\hbar k_T}\sum\limits_{j=1}^{N}(P_yA_j)
\dfrac{f_0({\bf P}+\dfrac{{\bf q}_j}{2})-f_0({\bf P}-\dfrac{{\bf q}_j}{2})}
{q_jP_x-\Omega_j},
$$
or
$$
f_1=-\dfrac{e}{c\hbar k_T}\sum\limits_{j=1}^{N}({\bf PA}_j)\dfrac{f_0({\bf P}+
\dfrac{{\bf q}_j}{2})-f_0({\bf P}-\dfrac{{\bf q}_j}{2})}{{\bf q}_j{\bf P}_j-\Omega_j}.
\eqno{(2.3)}
$$

\begin{center}
\item{}\section{The solution of Wigner equation in second approximation}
\end{center}

In the second approximation we search for the decision of Wigner equation (1.2) in
the form of (1.3), in which $f_2$ it is defined by equality (1.5). By means of a
method of small parameter we will substitute in the first two members of equation
(1.2), we substitute $f_2$ and in the third member of equation we substitute $f_1$.
In the fourth and fifth members of equation we substitute $f_0$. We receive the
following equation
$$
\dfrac{\partial f_2}{\partial t}+v_TP_x
\dfrac{\partial f_2}{\partial x}+
\dfrac{ie v_T}{c\hbar}\sum\limits_{j=1}^{N}(P_y A_j)
\Big[f_1\big(x,{\bf P}+\dfrac{{\bf q}_j}{2},t\big)-
f_1\big(x,{\bf P}-\dfrac{{\bf q}_j}{2},t\Big)\big]-
$$
$$
-\dfrac{ie^2}{mc^2\hbar}\sum\limits_{\substack{s,j=1\\j<s}}^{N}A_sA_j
\Big[f_0\big({\bf P}+\dfrac{{\bf q}_s+{\bf q}_j}{2}\big)-
f_0\big({\bf P}-\dfrac{{\bf q}_s+{\bf q}_j}{2}\big)\Big]=0.
\eqno{(1.2'')}
$$

Calculating the first two terms of equation
and using the first approximation (2.3), we obtain the following equation

$$
iv_Tk_T\sum\limits_{s,j=1}^{N}\Big[(q_s+q_j)P_x-(\Omega_s+\Omega_j)\Big]A_sA_j\xi_{s,j}=
$$
$$
=\dfrac{ie^2}{c^2m\hbar}\sum\limits_{j,s=1}^{N}P_y^2A_jA_s\Bigg[
\dfrac{f_0\Big({\bf P}+\dfrac{{\bf q_s+q_j}}{2}\Big)-
f_0\Big({\bf P}-\dfrac{{\bf q_s-q_j}}{2}\Big)}{q_s(P_x+q_j/2)-\Omega_s}-
$$
$$
-\dfrac{f_0\Big({\bf P}+\dfrac{{\bf q_s-q_j}}{2}\Big)-
f_0\Big({\bf P}-\dfrac{{\bf q_s+q_j}}{2}\Big)}{q_s(P_x-q_j/2)-\Omega_s}\Bigg]+
$$
$$
+\dfrac{ie^2}{2c^2m\hbar}\sum\limits_{j,s=1}^{N}A_jA_s\Big[f_0\Big({\bf P}+
\dfrac{{\bf q_j+q_s}}{2}\Big)-f_0\Big({\bf P}-\dfrac{{\bf q_j+q_s}}{2}\Big)\Big].
\eqno{(3.1)}
$$

Let us find the sum of differences of the third term from the equation (3.1)
$$
-\dfrac{iev_{T}}{c\hbar}\left[\left(\textbf{P}\textbf{A}_{1} \right)\left[ f_{1}
\left(\textbf{P}+\frac{\textbf{q}_{1}}{2}\right)  -f_{1}\left(\textbf{P}-
\frac{\textbf{q}_{1}}{2}\right) \right]+\cdots+\right.
$$
$$
\left.+\left[ f_{1}\left(\textbf{P}+\frac{\textbf{q}_{N}}{2}\right)  -f_{1}
\left(\textbf{P}-\frac{\textbf{q}_{N}}{2}\right) \right]   \right] =
$$
$$
=\dfrac{ie^{2}}{c^{2}m\hbar}\left\{ \sum\limits_{j,s=1}^{N}
\left(\textbf{P}\textbf{A}_j \right)\left(\textbf{P}\textbf{A}_s\right)
\left[\dfrac{f_{0}\left(\textbf{P}+\frac{\textbf{q}_{j}+\textbf{q}_{s}}{2} \right)-
f_{0}\left(\textbf{P}-\frac{\textbf{q}_{j}+\textbf{q}_{s}}{2} \right) }
{\textbf{P}\textbf{q}_s-\Omega_{s}+\frac{\textbf{q}_j\textbf{q}_{s}}{2}} \right]-\right.
$$
$$
-\left.
\left[\dfrac{f_{0}\left(\textbf{P}-\frac{\textbf{q}_{j}-\textbf{q}_{s}}{2} \right)-
f_{0}\left(\textbf{P}-\frac{\textbf{q}_{j}+\textbf{q}_{s}}{2} \right) }
{\textbf{P}\textbf{q}_s-\Omega_{s}+\frac{\textbf{q}_j\textbf{q}_{s}}{2}} \right]+\right.
$$
$$
\left.
+\left[\dfrac{f_{0}\left(\textbf{P}+\frac{\textbf{q}_{j}+\textbf{q}_{s}}{2} \right)-
f_{0}\left(\textbf{P}-\frac{\textbf{q}_{j}-\textbf{q}_{s}}{2} \right) }
{\textbf{P}\textbf{q}_j-\Omega_{j}+\frac{\textbf{q}_j\textbf{q}_{s}}{2}} \right]-
\right.
$$
$$
\left.
-\left[\dfrac{f_{0}\left(\textbf{P}+\frac{\textbf{q}_{j}-\textbf{q}_{s}}{2} \right)-
f_{0}\left(\textbf{P}-\frac{\textbf{q}_{j}-\textbf{q}_{s}}{2} \right) }
{\textbf{P}\textbf{q}_j-\Omega_{j}+\frac{\textbf{q}_j\textbf{q}_{s}}{2}} \right]
\right\}.
$$

The composed equalities (1.5) are equal
$$
\textbf{A}_{j}\textbf{A}_{s}\xi_{j,s}=\dfrac{e^{2}}{2c^{2}p_{T}^{2}}\times
\dfrac{(\textbf{P}\textbf{A}_{j})(\textbf{P}\textbf{A}_{s})}{\textbf{P}\dfrac{\textbf{q}_j+
\textbf{q}_s}{2} -\dfrac{\Omega_{j}+\Omega_{s}}{2}}\times
 $$
 $$
 \times \left\{ \left[\dfrac{f_{0}\left(\textbf{P}+\frac{\textbf{q}_{j}+\textbf{q}_{s}}{2}
 \right)-f_{0}\left(\textbf{P}-\frac{\textbf{q}_{j}+\textbf{q}_{s}}{2}\right)}
 {\textbf{P}\textbf{q}_s-\Omega_{s}+\frac{\textbf{q}_j\textbf{q}_{s}}{2}} \right]\right.-
 $$
 $$
 \left.
 -\left[\dfrac{f_{0}\left(\textbf{P}-\frac{\textbf{q}_{j}-\textbf{q}_{s}}{2}\right)-
 f_{0}\left(\textbf{P}-\frac{\textbf{q}_{j}+\textbf{q}_{s}}{2}\right)}
 {\textbf{P}\textbf{q}_s-\Omega_{s}+\frac{\textbf{q}_j\textbf{q}_{s}}{2}} \right]+\right.
 $$
 $$
 \left.
 +\left[\dfrac{f_{0}\left(\textbf{P}+\frac{\textbf{q}_{j}+\textbf{q}_{s}}{2}\right)-
 f_{0}\left(\textbf{P}-\frac{\textbf{q}_{j}-\textbf{q}_{s}}{2}\right)}
 {\textbf{P}\textbf{q}_j-\Omega_{j}-\frac{\textbf{q}_j\textbf{q}_{s}}{2}} \right]-\right.
 $$
 $$
 \left.
 -\left[\dfrac{f_{0}\left(\textbf{P}+\frac{\textbf{q}_{j}-\textbf{q}_{s}}{2}\right)-
 f_{0}\left(\textbf{P}-\frac{\textbf{q}_{j}-\textbf{q}_{s}}{2} \right)}
 {\textbf{P}\textbf{q}_j-\Omega_{j}-\frac{\textbf{q}_j\textbf{q}_{s}}{2}} \right]\right\}+
\dfrac{e^{2}\textbf{A}_{j}\textbf{A}_{s}}{2c^{2}p_{T}^{2}}\times
$$
$$\times\dfrac{f_{0}
\left(\textbf{P}+\dfrac{\textbf{q}_{s}+\textbf{q}_{j}}{2} \right)-f_{0}
\left(\textbf{P}-\dfrac{\textbf{q}_{s}+\textbf{q}_{j}}{2} \right)}
{\textbf{P}\dfrac{\textbf{q}_j+\textbf{q}_s}{2}-\dfrac{\Omega_j+\Omega_s}{2}},\quad
s,j=1,2,\cdots,N.
$$
and

$$
\textbf{A}^2_b\psi_b=\dfrac{e^2}{2c^2p^2_T}\left[\left(\textbf{P}\textbf{A}_b \right)^2
\dfrac{f_0\left(\textbf{P}+\textbf{q}_b\right)-f_0\left(\textbf{P}-\textbf{q}_b\right)}
{\textbf{P}\textbf{q}_b-\Omega_b+\dfrac{\textbf{q}_b^2}{2}}+\right.
$$
$$
\left.
+\frac{\textbf{A}_b^2}{2}\left(f_0(\textbf{P}+\textbf{q}_b)-f_0(\textbf{P}-
\textbf{q}_b)\right)\right]\dfrac{1}{\textbf{P}\textbf{q}_b-\Omega_b} \quad
b=1,2,\cdots,N
\eqno{(3.2)}
$$

Thus, the decision of Wigner equation is constructed and in the second
approximation. It is defined by equalities (1.3)--(1.5), in which functions
$\xi_{s,j} (s,j=1,2,\cdots,N)$ and $ \psi_{b}(b=1,2,\cdots,N) $ are defined equality
(3.2).

\begin{center}
  \item{}\section{The electric current in quantum plasma}
\end{center}

The density of electric current according to his definition is equal
$$
{\bf j}=e\int f {\bf v}\dfrac{2d^3p}{(2\pi \hbar)^3}.
$$

In work  \cite{Lat1} it is shown, that in zeroth approximation electric current in
quantum plasmas equals to zero
$$
{\bf j}^{(0)}=e\int f_0(P) {\bf v}\dfrac{2d^3p}{(2\pi \hbar)^3}=0.
$$

Therefore the density of electric current in quantum plasmas is equal
$$
{\bf j}=\dfrac{2ep_T^3v_T}{(2\pi\hbar)^3}\int (f_1+f_2)
\Big({\bf P}-\dfrac{e({\bf A}_1+{\bf A}_2+\cdots+\textbf{A}_N)}{mcv_T}\Big)d^3P.
\eqno{(4.1)}
$$

Equality (4.1) can be presented in the form
$$
{\bf j}={\bf j}^{\rm linear}+{\bf j}^{\rm quadr}.
$$

Here
$$
{\bf j}^{\rm linear}=\dfrac{2ep_T^3v_T}{(2\pi\hbar)^3}\int f_1
{\bf P}d^3P,
\eqno{(4.2)}
$$
$$
{\bf j}^{\rm quadr}=\dfrac{2ep_T^3v_T}{(2\pi\hbar)^3}\int \Big[f_2
{\bf P}-\dfrac{e({\bf A}_1+{\bf A}_2+\cdots+\textbf{A}_N)}{cp_T}f_1\Big]d^3P.
\eqno{(4.3)}
$$

Electric current in quantum plasma is the sum two composed, linear and square.
Linear composed there is density of the current directed along the vector capacity
of the electromagnetic field (i.e. along vector of tension of the field).
It consists of the members proportional to the first degree of vector potentials.
Square composed there is density of current, orthogonal current of linear density.
It is directed along wave vector. Square composed consists of the members
proportional to square of vector potentials of current and their work.

Let us present linear part of density of current (4.2) in an explicit form
$$
{\bf j}^{\rm linear}=-\dfrac{2ep_T^3v_T}{(2\pi\hbar)^3c\hbar k_T}
\sum\limits_{j=1}^{N}\int {\bf P(PA)}_j \dfrac{f_0\Big({\bf P}+\dfrac{{\bf q}_j}{2}\Big)-
f_0\Big({\bf P}-\dfrac{{\bf q}_j}{2}\Big)}{{\bf Pq}_j-\Omega_j}d^3P.
$$

This vector expression has one nonzero to component
$$
{\bf j}^{\rm linear}=j_y(0,1,0),
$$
where
$$
j_y=-\dfrac{2ep_T^3}{(2\pi\hbar)^3cm}
\sum\limits_{j=1}^{N}A_j \times
$$
$$
\times\int
\dfrac{f_0\Big(P_x+\dfrac{q_j}{2},P_y,P_z\Big)-
f_0\Big(P_x-\dfrac{q_j}{2},P_y,P_z\Big)}{q_jP_x-\Omega_j}P_y^2d^3P.
\eqno{(4.4)}
$$

Here designation is entered
$$
f_0\Big(P_x\pm\dfrac{q_j}{2},P_y,P_z\Big)=\dfrac{1}{1+
\exp\Big[(P_x\pm\dfrac{q_j}{2})^2+P_y^2+P_z^2-\alpha\Big]}.
$$

Let us use further shorter designation
$$
f_0\Big(P_x\pm\dfrac{q_j}{2}\Big)\equiv
f_0\Big(P_x\pm\dfrac{q_j}{2},P_y,P_z\Big).
$$

Let us find numerical density the concentration of particles of plasma answering to
the distribution of Fermi---Dirac
$$
N_{0}=\int f_0(P)\dfrac{2d^3p}{(2\pi\hbar)^3}=
\dfrac{8\pi p_T^3}{(2\pi\hbar)^3}\int\limits_{0}^{\infty}
\dfrac{e^{\alpha-P^2}P^2dP}{1+e^{\alpha-P^2}}=
\dfrac{k_T^3}{2\pi^2}l_0(\alpha),
$$
where
$$
l_0(\alpha)=\int\limits_{0}^{\infty}\ln(1+e^{\alpha-\tau^2})d\tau.
$$

Let us enter a change of variables in (4.2) and we will enter plasma (Langmuir)
frequency
$$
\omega_p=\sqrt{\dfrac{4\pi e^2 N}{m}}.
$$
Then we use communication between the numerical density of particles plasmas
(concentration), thermal wave number of electrons and them chemical potential
$$
N_{0}=\dfrac{1}{2\pi^2}k_T^3l_0(\alpha).
$$

As a result we receive that expression for current (4.4) is equal
$$
j_y=-\dfrac{\omega_p^2}{8\pi^2c}\sum\limits_{j=1}^{N}A_j
\times $$$$ \times
\int
\Big(\dfrac{1}{q_jP_x-\Omega_j-q_j^2/2}-
\dfrac{1}{q_jP_x-\Omega_j+q_j^2/2}\Big)f_0(P)P_y^2d^3P,
$$
or
$$
j_y=i\dfrac{\omega_p^2}{8\pi^2}\sum\limits_{j=1}^{N}
\dfrac{E_jq_j^2}{\omega_j}\int
\dfrac{f_0(P)P_y^2d^3P}{(q_jP_x-\Omega_j)^2-q_j^4/4}.
$$

This expression of density of transversal current comes down to double  integral
$$
j_y=\dfrac{i\Omega_p^2 k_Tv_T}{8\pi}\sum\limits_{j=1}^{N}
\dfrac{E_jq_j^2}{\Omega_j}\int\limits_{0}^{\infty}
\dfrac{P^4dP}{1+e^{P^2-\alpha}}\int\limits_{-1}^{1}
\dfrac{(1-\mu^2)d\mu}{(q_jP\mu-\Omega_j)^2-q_j^4/4}.
$$

Here
$$
\Omega_p=\dfrac{\omega_p}{k_Tv_T}=\dfrac{\hbar\omega_p}{mv_T^2}
$$
is the dimensionless  plasma frequency.

\begin{center}
  \item{}\section{The longitudinal current in quantum plasma}
\end{center}

We should note that the integral from an addend in (4.3) is equal to zero. Therefore
the longitudinal current in quantum plasma generated by N electromagnetic fields is
equal
$$
{\bf j}^{\rm quadr} \equiv {\bf j}^{\rm long}=
\dfrac{2ep_T^3v_T}{(2\pi\hbar)^3}\int f_2{\bf P}d^3P.
\eqno{(5.1)}
$$

Thus, longitudinal current is defined only by the second approach of function of
distribution.

Vector equality (5.1) has only one nonzero to component ${\bf j}^{\rm
long}=j_x(1,0,0)$, where
$$
j_x=\dfrac{2ep_T^3v_T}{(2\pi\hbar)^3}\int f_2{P_x}d^3P.
\eqno{(5.2)}
$$

Longitudinal current (5.2) we will present the sums of composed in the form
$$
j_x=\sum\limits_{b=1}^{N}j_{b}+\sum\limits_{\substack{j,s=1\\s<j}}^{N}j_{j,s}.
\eqno{(5.3)}
$$
Here
$$
j_{b}=A_{b}^{2}\dfrac{2ep_{T}^{3}v_{T}}{(2\pi\hbar)^{3}}\int P_x\psi_{b} d^3P,\quad
(b=1,2,\cdots,N),
\eqno{(5.4)}
$$
$$
j_{j,s}=
A_{j}A_{s}\dfrac{2ep_{T}^{3}v_{T}}{(2\pi\hbar)^{3}}\int P_x\xi_{j,s} d^3P\quad
(j,s=1,2,\cdots,N).
\eqno{(5.5)}
$$

From equality (5.3) follows, that the longitudinal current represents the sum of two
components. Currents from the first sum $j_b $ are generated by the corresponding
vector potentials of electromagnetic fields. Their quantities are proportional to
squares of these vector potentials. The second sum of currents which we name
"crossed"\,, are generated by interaction of electromagnetic fields among themselves
and are proportional to product of quantities of their vector potentials.

We present the formula (5.4) in the explicit form
$$
j_b=A_b^2\dfrac{e^3p_Tv_T}{(2\pi\hbar)^3c^2}\int \Bigg[\Big(\dfrac{f_0(P_x+q_b)-f_0(P)}
{q_bP_x-\Omega_b+q_b^2/2}-\dfrac{f_0(P)-f_0(P_x-q_b)}{q_bP_x-\Omega_b-q_b^2/2}\Big)P_y^2+
$$
$$
+\dfrac{f_0(P_x+q_b)-f_0(P_x-q_b)}{2}\Bigg]
\dfrac{P_xd^3P}{q_bP_x-\Omega_b},\quad (b=1,2,\cdots,N).
\eqno{(5.6)}
$$

Let us transform the expression facing integral in the previous formula
$$
C=\dfrac{e^3p_Tv_T}{(2\pi\hbar)^3c^2}A_b^2=\dfrac{e^3k_T^3v_T}{8\pi^3c^2p_T^2}A_b^2.
$$
We will use communication between concentration (numerical density), thermal wave
number and chemical potential
$$
N_0=\dfrac{1}{2\pi^2}k_T^3l_0(\alpha), \quad
l_0(\alpha)=\int\limits_{0}^{\infty}\ln(1+e^{\alpha-\tau^2})d\tau.
$$
Then
$$
C=A_b^2\dfrac{2\pi^2e^3N_0v_T}{8\pi^3c^2p_T^2l_0(\alpha)}=
A_b^2\dfrac{e\omega_p^2}{16\pi^2c^2p_Tl_0(\alpha)}=
$$
$$
=-E_b^2\dfrac{e\omega_p^2}{16\pi^2l_0(\alpha)p_T\omega_b}=
-E_b^2\dfrac{e\Omega_p^2}{16\pi^2l_0(\alpha)p_T\Omega_b}.
$$
Here dimensionless plasma frequencies are entered
$$
\Omega_p=\dfrac{\omega_p}{k_Tv_T}, \qquad
\Omega_j=\dfrac{\omega_j}{k_Tv_T}\qquad (j=1,2,\cdots,N).
$$

We introduce the longitudinal-transversal conductivity $\sigma_{l,tr}$,
$$
\sigma_{l,tr}=\dfrac{e\hbar}{p_T^2}
\Big(\dfrac{\hbar \omega_p}{mv_T^2}\Big)^2=
\dfrac{e}{k_Tp_T}\Big(\dfrac{\omega_p}{k_Tv_T}\Big)^2=
\dfrac{e\Omega_p^2}{p_Tk_T}.
$$
Then we have
$$
C=-\dfrac{E_b^2
\sigma_{l,tr}k_b}{16\pi^2l_0(\alpha)\Omega_b^2q_b}, \quad
(b=1,2,\cdots,N).
$$

Now the formula (5.6)  can be presented as
$$
j_b=J_b\sigma_{l,tr}k_bE_b^2.
\eqno{(5.7)}
$$

In (5.7) $J_b$ is dimensionless current densities,
$$
J_b=-\dfrac{1}{16 \pi^2 l_0(\alpha)q_b\Omega_b}
\int\Bigg[\Big(
\dfrac{f_0(P_x+q_b)-f_0(P)}{q_bP_x-\Omega_b+q_b^2/2}-
\dfrac{f_0(P)-f_0(P_x-q_b)}{q_bP_x-\Omega_b-q_b^2/2}\Big)P_y^2+
$$
$$
+\dfrac{f_0(P_x+q_b)-f_0(P_x-q_b)}{2}\Bigg]
\dfrac{P_xd^3P}{q_bP_x-\Omega_b}.
\eqno{(5.8)}
$$

In (5.8) we will reduce integral to one-dimensional. For this purpose it will be
necessary for us following equalities. Let us calculate internal
integrals in $(P_y,P_z)$ passing to polar coordinates
$$
\int f_0(P_x\pm q_b,P_y,P_z)P_y^2dP_ydP_z=
\int\limits_{0}^{2\pi}\int\limits_{0}^{\infty}
\dfrac{\cos^2\varphi \rho^3 d\varphi d\rho}
{1+e^{(P_x\pm q_b)^2+\rho^2-\alpha}}=
$$
$$
=\pi\int\limits_{0}^{\infty}
\dfrac{\rho^3d\rho}{1+e^{(P_x\pm q_b)^2+\rho^2-\alpha}}=
\pi \int\limits_{0}^{\infty}
\rho\ln(1+e^{-(P_x\pm q_b)^2-\rho^2+\alpha})d\rho,
$$
where
$$
\rho=\sqrt{P_y^2+P_z^2}.
$$

Similarly 
$$
\int f_0(P)P_y^2dP_ydP_z=\int\limits_{0}^{2\pi}\int\limits_{0}^{\infty}
\dfrac{\cos^2\varphi \rho^3 d\varphi d\rho}{1+e^{P_x^2+\rho^2-
\alpha}}=
$$
$$
=\pi \int\limits_{0}^{\infty}
\rho\ln(1+e^{-P_x^2-\rho^2+\alpha})d\rho,
$$
$$
\int f_0(P\pm q_b)dP_ydP_z=\int\limits_{0}^{2\pi}\int\limits_{0}^{\infty}
\dfrac{\rho d\varphi d\rho}{1+e^{(P_x\pm q_b)^2+\rho^2-\alpha}}=
$$
$$
=2\pi\int\limits_{0}^{\infty}
\dfrac{e^{\alpha-(P_x\pm q_b)^2-\rho^2}}
{1+e^{\alpha-(P_x\pm q_b)^2-\rho^2}}\rho d\rho=
\pi \ln(1+e^{\alpha-(P_x\pm q_b)^2}).
$$
Therefore
$$
\int\limits_{-\infty}^{\infty}\int\limits_{-\infty}^{\infty}
f_0(P)dP_ydP_z=\pi\ln(1+e^{\alpha-P_x^2}).
$$

Let us introduce the following notation
$$
l(P_x\pm q)=\int\limits_{0}^{\infty}\rho\ln(1+
e^{-(P_x\pm q)^2-\rho^2+\alpha})d\rho,
$$
$$
l(P_x)=\int\limits_{0}^{\infty}\rho\ln(1+
e^{-P_x^2-\rho^2+\alpha})d\rho.
$$

The integral from an addend from (5.8) is equal
$$
\dfrac{1}{2}\int
\dfrac{f_0(P_x+q_b)-f_0(P_x-q_b)}{q_bP_x-\Omega_b}P_xd^3P=$$$$=
\dfrac{\Omega_b}{2q_b}\int\dfrac{f_0(P_x+q_b)-f_0(P_x-q_b)}
{q_bP_x-\Omega_b}d^3P=
$$
$$
=\dfrac{\pi \Omega_b}{2q_b}\int\limits_{-\infty}^{\infty}
\ln\dfrac{1+e^{\alpha-(\tau+q_b)^2}}{1+e^{\alpha-(\tau-q_b)^2}}
\dfrac{d\tau}{q_b\tau-\Omega_b}=
$$
$$
=\dfrac{\pi \Omega_b}{2q_b}\int\limits_{-\infty}^{\infty}
\ln(1+e^{\alpha-\tau^2})\Big(\dfrac{1}{q_b\tau-\Omega_b-q_b^2}-
\dfrac{1}{q_b\tau-\Omega_b+q_b^2}\Big)d\tau=
$$
$$
=\pi q_b\Omega_b\int\limits_{-\infty}^{\infty}
\dfrac{\ln(1+e^{\alpha-\tau^2})d\tau}{(q_b\tau-\Omega_b)^2-q_b^4}=
\dfrac{\pi \Omega_b}{q_b}\int\limits_{-\infty}^{\infty}
\dfrac{\ln(1+e^{\alpha-\tau^2})d\tau}{(\tau-\Omega_b/q_b)^2-q_b^2}.
$$

Let us calculate integral from the first item. We have
$$
\int\Big(\dfrac{f_0(P_x+q_b)-f_0(P)}{q_bP_x-\Omega_b+q_b^2/2}-
\dfrac{f_0(P)-f_0(P_x-q_b)}{q_bP_x-\Omega_b-q_b^2/2}\Big)
\dfrac{P_y^2P_xd^3P}{q_bP_x-\Omega_b}=
$$
$$
=\pi \int\limits_{-\infty}^{\infty}
\Big(\dfrac{L(P_x+q_b,P_x)}{q_jP_x-\Omega_b+q_b^2/2}+
\dfrac{L(P_x-q_b,P_x)}{q_bP_x-\Omega_b-q_b^2/2}\Big)
\dfrac{P_xdP_x}{q_bP_x-\Omega_b}.
$$
Here
$$
L(P_x\pm q_b,P_x)=l(P_x\pm q_b)-l(P_x)=\int\limits_{0}^{\infty}
\rho\ln \dfrac{1+e^{\alpha-(P_x\pm q_b)^2-\rho^2}}
{1+e^{\alpha-P_x^2-\rho^2}}d\rho.
$$

We will transform the considered integral as follows

$$
\int\limits_{-\infty}^{\infty}
\Big(\dfrac{L(\tau+q_b,\tau)}{q_b\tau-\Omega_b+q_b^2/2}+
\dfrac{L(\tau-q_b,\tau)}{q_b\tau-\Omega_b-q_b^2/2}\Big)
\dfrac{\tau d\tau}{q_b\tau-\Omega_b}=
$$
$$
=\int\limits_{-\infty}^{\infty}
\Big(\dfrac{\tau-q_b/2}{q_b\tau-\Omega_b-q_b^2/2}-
\dfrac{\tau+q_b/2,\tau)}{q_b\tau-\Omega_b+q_b^2/2}\Big)
\dfrac{L(\tau+q_b/2,\tau-q_b/2)}{q_b\tau-\Omega_b}d\tau=
$$
$$
=q_b\Omega_b\int\limits_{-\infty}^{\infty}
\dfrac{L(\tau+q_b/2,\tau-q_b/2)d\tau}{(q_b\tau-\Omega_b)
[(q_b\tau-\Omega_b)^2-q_b^4/4]}d\tau.
$$

The finally dimensionless current density is equal
$$
J_b=-\dfrac{1}{16\pi l_0(\alpha)\Omega_b}
\int\limits_{-\infty}^{\infty}\Bigg[
\dfrac{L(\tau+q_b/2,\tau-q_b/2)}{(q_b\tau-\Omega_b)
(q_b\tau-\Omega_b)^2-q_b^4/4}+
$$
$$+
\dfrac{\ln(1+e^{\alpha-\tau^2})}{(q_b\tau-\Omega_b)^2-q_b^4}\Bigg]d\tau.
$$

\begin{center}
  \item{}\section{The crossed current}
\end{center}

We rewrite the formula for calculation of crossed currents in the explicit form
$$
\sum\limits_{\substack{s,j=1\\j<s}}^{N}j_{j,s}=
\dfrac{e^3p_T^3v_T}{(2\pi\hbar)^3c^2p_T^2}\sum\limits_{\substack{s,j=1\\j<s}}^{N}A_jA_s\int
\Bigg[\dfrac{f_0(P_x+q^+)-f_0(P_x-q^-)}{q_jP_x-\Omega_j+q_jq_s/2}-
$$
$$
-\dfrac{f_0(P_x+q^-)-f_0(P_x-q^+)}{q_jP_x-\Omega_j-q_jq_s/2}+
\dfrac{f_0(P_x+q^+)-f_0(P_x+q^-)}{q_sP_x-\Omega_s+q_jq_s/2}-
$$
$$
-\dfrac{f_0(P_x-q^-)-f_0(P_x-q^+)}{q_sP_x-\Omega_s-q_jq_s/2}\Bigg]
\dfrac{P_y^2P_xd^3P}{qP_x-\Omega}+
$$
$$
+\dfrac{e^3p_T^3v_T}{(2\pi\hbar)^3c^2p_T^2}\sum\limits_{\substack{s,j=1\\j<s}}^{N}A_jA_s\int
\dfrac{f_0(P_x+q)-f_0(P_x-q)}{qP_x-\Omega}P_xd^3P.
$$

We rewrite this equation

$$
\sum\limits_{\substack{s,j=1\\j<s}}^{N}j_{j,s}=\dfrac{e^3p_Tv_T}{(2\pi\hbar)^3c^2}
\sum\limits_{\substack{s,j=1\\j<s}}^{N} A_jA_s(J_1-J_2+J_3-J_4+J_5).
\eqno{(6.1)}
$$
Here
$$
J_1=\int\dfrac{f_0(P_x+q^+)-f_0(P_x-q^-)}{q_jP_x-\Omega_j+q_jq_s/2}
\dfrac{P_y^2P_xd^3P}{qP_x-\Omega},
$$
$$
J_2=\int\dfrac{f_0(P_x+q^-)-f_0(P_x-q^+)}{q_jP_x-\Omega_j-q_jq_s/2}
\dfrac{P_y^2P_xd^3P}{qP_x-\Omega},
$$
$$
J_3=\int\dfrac{f_0(P_x+q^+)-f_0(P_x+q^-)}{q_sP_x-\Omega_s+q_jq_s/2}
\dfrac{P_y^2P_xd^3P}{qP_x-\Omega},
$$
$$
J_4=\int\dfrac{f_0(P_x-q^-)-f_0(P_x-q^+)}{q_sP_x-\Omega_s-q_jq_s/2}
\dfrac{P_y^2P_xd^3P}{qP_x-\Omega},
$$
$$
J_5=\int\dfrac{f_0(P_x+q)-f_0(P_x-q)}{qP_x-\Omega}P_xd^3P.
$$

Here
$$
q=q^+=\dfrac{q_j+q_s}{2},\quad q^-=\dfrac{q_j-q_s}{2}, \quad
\Omega=\Omega^+=\dfrac{\Omega_j+\Omega_s}{2}.
$$

For this purpose it will be necessary for us following equalities. Let us calculate
internal integrals in $(P_y,P_z)$ passing to polar coordinates
$$
\int f_0(P_x\pm q^\pm,P_y,P_z)P_y^2dP_ydP_z=
\int\limits_{0}^{2\pi}\int\limits_{0}^{\infty}
\dfrac{\cos^2\varphi \rho^3 d\varphi d\rho}
{1+e^{(P_x\pm q^\pm)^2+\rho^2-\alpha}}=
$$
$$
=\pi\int\limits_{0}^{\infty}
\dfrac{\rho^3d\rho}{1+e^{(P_x\pm q^\pm)^2+\rho^2-\alpha}}=
\pi \int\limits_{0}^{\infty}
\rho\ln(1+e^{-(P_x\pm q^\pm)^2-\rho^2+\alpha})d\rho,
$$
where
$$
\rho=\sqrt{P_y^2+P_z^2}.
$$

Similarly
$$
\int f_0(P)P_y^2dP_ydP_z=$$$$=\int\limits_{0}^{2\pi}\int\limits_{0}^{\infty}
\dfrac{\cos^2\varphi \rho^3 d\varphi d\rho}{1+e^{P_x^2+\rho^2-
\alpha}}=
\pi \int\limits_{0}^{\infty}
\rho\ln(1+e^{-P_x^2-\rho^2+\alpha})d\rho,
$$
$$
\int f_0(P\pm q)dP_ydP_z=\int\limits_{0}^{2\pi}\int\limits_{0}^{\infty}
\dfrac{\rho d\varphi d\rho}{1+e^{(P_x\pm q)^2+\rho^2-\alpha}}=
$$
$$
=2\pi\int\limits_{0}^{\infty}
\dfrac{e^{\alpha-(P_x\pm q)^2-\rho^2}}
{1+e^{\alpha-(P_x\pm q)^2-\rho^2}}\rho d\rho=
\pi \ln(1+e^{\alpha-(P_x\pm q)^2}).
$$

Let us introduce the following notation
$$
l(P_x\pm q)=\int\limits_{0}^{\infty}\rho\ln(1+
e^{-(P_x\pm q)^2-\rho^2+\alpha})d\rho,
$$
$$
l(P_x)=\int\limits_{0}^{\infty}\rho\ln(1+
e^{-P_x^2-\rho^2+\alpha})d\rho.
$$

After long transformations $J_1,\cdots,J_5$ can be reduced to one-dimensional
$$
J_1=\pi\int\limits_{-\infty}^{\infty}\dfrac{l(\tau+q^+)-l(\tau-q^-)}
{q_j\tau-\Omega_j+q_jq_s/2}\cdot \dfrac{\tau d\tau}{q\tau-\Omega},
$$
$$
J_2=\pi\int\limits_{-\infty}^{\infty}\dfrac{l(\tau+q^-)-l(\tau-q^+)}
{q_j\tau-\Omega_j-q_jq_s/2}\cdot \dfrac{\tau d\tau}{q\tau-\Omega},
$$
$$
J_3=\pi\int\limits_{-\infty}^{\infty}\dfrac{l(\tau+q^+)-l(\tau+q^-)}
{q_s\tau-\Omega_s+q_jq_s/2}\cdot \dfrac{\tau d\tau}{q\tau-\Omega},
$$
$$
J_4=\pi\int\limits_{-\infty}^{\infty}\dfrac{l(\tau-q^-)-l(\tau-q^+)}
{q_s\tau-\Omega_s-q_jq_s/2}\cdot \dfrac{\tau d\tau}{q\tau-\Omega},
$$
$$
J_5=2\pi q^+\Omega\int\limits_{-\infty}^{\infty}
\dfrac{\ln(1+e^{\alpha-\tau^2})d\tau}{(q^+\tau-\Omega)^2-{q^+}^4}.
$$

Numerators are respectively equal in integrals $J_1,\cdots,J_4$
$$
l(\tau+q^+)-l(\tau-q^-)=\int\limits_{0}^{\infty}
\rho\ln\dfrac{1+e^{\alpha-(\tau+q^+)^2-\rho^2}}
{1+e^{\alpha-(\tau-q^-)^2-\rho^2}}d\rho,
$$
$$
l(\tau+q^-)-l(\tau-q^+)=\int\limits_{0}^{\infty}
\rho\ln\dfrac{1+e^{\alpha-(\tau+q^-)^2-\rho^2}}
{1+e^{\alpha-(\tau-q^+)^2-\rho^2}}d\rho,
$$
$$
l(\tau+q^+)-l(\tau+q^-)=\int\limits_{0}^{\infty}
\rho\ln\dfrac{1+e^{\alpha-(\tau+q^+)^2-\rho^2}}
{1+e^{\alpha-(\tau+q^-)^2-\rho^2}}d\rho,
$$
$$
l(\tau-q^-)-l(\tau-q^+)=\int\limits_{0}^{\infty}
\rho\ln\dfrac{1+e^{\alpha-(\tau-q^-)^2-\rho^2}}
{1+e^{\alpha-(\tau-q^+)^2-\rho^2}}d\rho.
$$

Let us enter into a formula (6.1) a plasma frequency. Now the formula (6.1)  can be
presented as
$$
\sum\limits_{\substack{s,j=1\\j<s}}^{N}j_{j,s}=
\dfrac{e\omega_p^2}{16\pi^2 c^2p_Tl_0(\alpha)}\sum\limits_{\substack{s,j=1\\j<s}}^{N}
A_jA_s(J_1-J_2+J_3-J_4+J_5).
\eqno{(6.2)}
$$

In a formula (6.2) we will pass from sizes of vector potentials
electromagnetic fields to strengths of electric fields
$$
\sum\limits_{\substack{s,j=1\\j<s}}^{N}j_{j,s}=$$$$=
-\dfrac{e\Omega_p^2}{16\pi^2 l_0(\alpha)\Omega_j\Omega_sp_T}
\sum\limits_{\substack{s,j=1\\j<s}}^{N}E_jE_s
(J_1-J_2+J_3-J_4+J_5).
\eqno{(6.3)}
$$

We introduce again the longitudinal-transversal conductivity $\sigma_{l,tr}$
$$
\sigma_{l,tr}=\dfrac{e\hbar}{p_T^2}
\Big(\dfrac{\hbar \omega_p}{mv_T^2}\Big)^2=
\dfrac{e}{k_Tp_T}\Big(\dfrac{\omega_p}{k_Tv_T}\Big)^2=
\dfrac{e\Omega_p^2}{p_Tk_T}.
$$

Now the formula (6.3) can be presented
$$
\sum\limits_{\substack{s,j=1\\j<s}}^{N}j_{j,s}=$$$$=-\dfrac{\sigma_{l,tr}}
{16\pi^2 l_0(\alpha)}
\sum\limits_{\substack{s,j=1\\j<s}}^{N}\dfrac{E_jE_s(k_j+k_s)}
{\Omega_j\Omega_s(q_j+q_s)}(J_1-J_2+J_3-J_4+J_5).
\eqno{(6.4)}
$$

We will write  a formula (6.4) in the form
$$
\sum\limits_{\substack{s,j=1\\j<s}}^{N}j_{j,s}=
\sum\limits_{\substack{s,j=1\\j<s}}^{N}J_{j,s}\sigma_{l,tr}E_jE_s(k_j+k_s).
\eqno{(6.5)}
$$

In a formula (6.5) $J_{j,s}$ is the dimensionless part of density cross current
$$
\sum\limits_{\substack{s,j=1\\j<s}}^{N}J_{j,s}=$$$$=
-\dfrac{1}{16\pi^2 l_0(\alpha)}\sum\limits_{\substack{s,j=1\\j<s}}^{N}
\dfrac{1}{\Omega_j\Omega_s(q_j+q_s)}
(J_1-J_2+J_3-J_4+J_5).
$$

Thus, a longitudinal part of current is equal
$$
j_x=\sigma_{l,tr}\left[ \sum\limits_{b=1}^{N}E_b^2k_bJ_b+
\sum\limits_{\substack{s,j=1\\j<s}}^{N}+E_jE_s(k_j+k_s)J_{j,s}\right] .
\eqno{(6.6)}
$$

If to enter transversal fields
$$
\mathbf{E}_j^{\bf \rm tr}=\mathbf{E}_j-
\dfrac{\mathbf{k}_j({\bf E}_j{\bf k}_j)}{k_j^2}=
\mathbf{E}_j-\dfrac{\mathbf{q}_j({\bf E}_j{\bf q}_j)}{q_j^2},\quad
(j=1,2,\cdots,N).
$$
then equality (6.6) can be written down in an invarianty form
$$
{\bf j}^{\rm long}=\sigma_{l,tr}\left[ \sum\limits_{b=1}^{N}({\bf E}_b^{tr})^2{\bf k}_bJ_b+
\sum\limits_{\substack{s,j=1\\j<s}}^{N}{\bf E}_j^{tr}{\bf E}_s^{tr}
({\bf k}_j+{\bf k}_s)J_{j,s}\right] .
$$

\begin{center}
  \item{}\section{Small values of wave numbers}
\end{center}

In case of small values of wave numbers of size of the currents proportional to squares
of strengths of electric fields, are actually calculated in work  \cite{Lat7}
$$
j_m=-\dfrac{e}{8\pi
\omega_m}\Big(\dfrac{\omega_p}{\omega_m}\Big)^2k_mE_m^2=
-\dfrac{e\Omega_p^2}{8\pi \omega_m\Omega_m^2}k_mE_m^2=
$$
$$
=-\dfrac{\sigma_{l,tr}}{8\pi \Omega_m^3}k_mE_m^2,\quad q_m\to 0,(m=1,2,\cdots,N).
$$

Now we will consider the size of a cross current at small values of wave numbers. We
get the equation
$$
f_0(P_x+q)=f_0(P)-g(P)2P_xq+\cdots, \qquad (q\to 0).
$$
Here
$$
g(P)=\dfrac{e^{P^2-\alpha}}{(1+e^{P^2-\alpha})^2}=
\dfrac{e^{\alpha-P^2}}{(1+e^{\alpha-P^2})^2}.
$$
Let us notice that
$$
f_0(P+q)-f_0(P-q)=-4g(P)P_xq+\cdots,\qquad (q\to 0).
$$

Therefore, density of the cross current is equal
$$
\sum\limits_{\substack{s,j=1\\j<s}}^{N}j_{j,s}=$$$$=-\dfrac{\sigma_{l,tr}}{4\pi^2l_0(\alpha)}
\sum\limits_{\substack{s,j=1\\j<s}}^{N}E_jE_s(k_j+k_s)\dfrac{q}{(q_j+q_s)\Omega_j
\Omega_s\Omega}\int g(P)P_x^2d^3P.
$$

The integral is equal
$$
\int g(P)P_x^2d^3P=\dfrac{4\pi}{3}\int\limits_{0}^{\infty}
\dfrac{P^4e^{P^2-\alpha}dP}{(1+e^{P^2-\alpha})^2}=\pi l_0(\alpha).
$$

Therefore, the density of crossed currents equals
$$
\sum\limits_{\substack{j,s=1\\j<s}}^{N}j_{j,s}=-\dfrac{\sigma_{l,tr}}{4\pi^2l_0(\alpha)}
\sum\limits_{\substack{j,s=1\\j<s}}^{N}
E_jE_s(k_j+k_s)\dfrac{q}{(q_j+q_s)\Omega_j\Omega_s\Omega}.
$$

We receive at small values wave numbers for density of longitudinal current
$$
j_x=-\dfrac{\sigma_{l,tr}}{8\pi}\Big[\sum\limits_{b=1}^{N}E_b^2\dfrac{k_1}{\Omega_b^3}+
2\sum\limits_{\substack{s,j=1\\j<s}}^{N}E_jE_s\dfrac{k_j+k_s}
{\Omega_j\Omega_s(\Omega_j+\Omega_s)}\Big].
$$

We write this formula in a vector form
$$
{\bf j}^{\rm long}=
-\dfrac{\sigma_{l,tr}}{8\pi}
\Big[\sum\limits_{b=1}^{N}({\bf E}_b^{tr})^2{\bf k}_b\dfrac{b}{\Omega_b^3}+
2\sum\limits_{\substack{s,j=1\\j<s}}^{N}{\bf E}_j^{tr}{\bf E}_s^{tr}({\bf k}_j+{\bf k}_s)
\dfrac{1}{\Omega_j\Omega_s(\Omega_j+\Omega_s)}\Big].
$$

This formula in accuracy matches the corresponding formula from work \cite{Lat10}.
It means that at small values of wave numbers the size of density of longitudinal
current in classical and quantum plasma matches.

{\sc Remark.} At calculation of the singular integrals entering dimensionless parts
of density of longitudinal current it is necessary to use the known rule of Landau.

\begin{center}
  \item{}\section{Conclusions}
\end{center}

This article is the continuous of our works \cite{Lat9}--\cite{Lat14}.
In the present work the following problem is solved: in quantum plasma with arbitrary
degree of degeneration of electronic gas, propagate N electromag\-ne\-tic waves with collinear
wave vectors. The Wigner equation dares by a method consecutive, approximately
considering as small parameteres of one order of quantities of intensities corresponding
electric fields. Square-law decomposi\-tion of function of distribution is used.

It has appeared, that the account of nonlinearity of electromagnetic fields finds out
generating of an electric current, orthogonal electric field to a direction
(i.e. to a direction of a known classical transversal electric current).
The quantities of transversal and longitudinal electric currents are found.
The case of small values of wave numbers is considered. It turned out, that the
quantities of a longitudinal current in classical and quantum plasma coincides.

Further authors purpose to consider problems about fluctuations of plasma and about
skin-effect with use square-law on potential of decomposition of function of distribution.

\end{document}